# Understanding Health and Behavioral Trends of Successful Students through Machine Learning Models

Abigale Kim[1], Fateme Nikseresht[2], Janine M. Dutcher[1], Michael Tumminia[3], Daniella Villalba[1], Sheldon Cohen[1], Kasey Creswell[1], David Creswell[1], Anind K. Dey[4], Jennifer Mankoff[4] and Afsaneh Doryab[2]

[1] Carnegie Mellon University, Pittsburgh, USA,
[2] University of Virginia, Charlottesville, USA,
[3] University of Pittsburgh, Pittsburgh, USA,
[4] University of Washington, Seattle, USA
abigalek@andrew.cmu.edu, fn5an@virginia.edu, jdutcher@andrew.cmu.edu,
mjtumminia@pitt.edu, daniella.villalba@gmail.com, scohen@cmu.edu,
kasey@andrew.cmu.edu, creswell@cmu.edu, anind@uw.edu, jmankoff@uw.edu,
ad4ks@virginia.edu

**Abstract.** This study analyzes patterns of physical, mental, lifestyle, and personality factors in college students in different periods over the course of a semester and models their relationships with students' academic performance. The data analyzed was collected through smartphones and Fitbit. The use of machine learning models derived from the gathered data was employed to observe the extent of students' behavior associated with their GPA, lifestyle, physical health, mental health, and personality attributes. A mutual agreement method was used in which rather than looking at the accuracy of results, the model parameters and weights of features were used to find common behavioral trends. From the results of the model creation, it was determined that the most significant indicator of academic success defined as a higher GPA, was the places a student spent their time. Lifestyle and personality factors were deemed more significant than mental and physical factors. This study will provide insight into the impact of different factors and the timing of those factors on students' academic performance.

## 1    Introduction

College students often have a rigorous workload, resulting in difficulty balancing other aspects of their lives including health and wellness while being academically successful. The time spent on classes alone, attending, studying, and completing assignments for them can account for approximately forty hours a week. In addition to this, extracurricular activities and attending social events compose another significant portion of time. As a result, college students are often faced with the adversity of balancing health, sleep, social, and academic needs. Students who do not do so properly can have deteriorating academic performance or health status at an alarming rate. Studies correlating academic performance to health and behavioral factors can identify flaws and help alter the college education system. If health and lifestyle trends of students with high academic success are determined, these lifestyle habits can be reinforced among students to optimize their academic success.

Advances in mobile and wearable technology has made it possible to track and study human behavior in the wild. Smartphones are already equipped with light, proximity, and microphone sensors which can indicate the type of environment the user of the smartphone is in. They can also log SMS and calls records as well as locations the person spends time in to indicate the user's social activity level. Although numerous studies have used college students as the target population to understand health and lifestyle trends as well as to test and evaluate different technology, a few of them have focused on understanding the impact of lifestyle and behavior on health and academic success [1, 2, 3, 4, 5, 6, 7, 8]. To our knowledge, only the research conducted by Dartmouth College (i.e., StudentLife and SmartGPA [1, 3] uses data from mobile devices over the course of a semester to objectively assess behavioral trends of students and the relationship to their GPA. Other studies use questionnaires and self-reported data for their analyses.

While our study asks a similar question of understanding the behavioral trends associated with student success through mobile and wearable devices, we 1) use a different analysis method and 2) aim to investigate how behavioral trends may differ in a larger study population in a different university located in a different geographical area. We analyze data from smartphone and Fitbit devices of 138 first-year students at an American University during one long semester of 16 weeks. To extract the behavioral trends, we first divide the semester-long data into three academic periods, namely before midterms, during midterms, and during finals. This division is made for this data-set because of the specific academic calendar at the studied university. We then develop an analysis method that first builds a machine learning model of the students' data for each academic period using two different learning algorithms; Logistic Regression and Support Vector Machine. Then investigates the built models to extract significant behavioral factors from them identified via weight coefficients. Finally, we use a mutual agreement method that takes the weights obtained from the two learned models to estimate the final weight of each behavioral and health factor.

The results obtained from all three periods are compared to each other in the analyses to determine how changes in the four factors- physical, mental, lifestyle, and personality correlate to academic success. We hypothesize that all four types of factors will have a significant impact on a student's academic success. The following sections describe our approach and result in more details. We first discuss the results of existing work in observing each of the four factors followed by the description of data collection and processing. We then describe the analysis method and discuss our results compared to existing work.

## 2 Related Work

### 2.1 Physical and Mental Health

Mixed results have been found regarding the correlation of one's exercise routine to their grade point average. For instance, the SmartGPA study [1] found that students who had higher indoor mobility prior to the midterm compared to after the midterm had higher GPAs. Lower mobility may be indicative of focused study later in the semester, which may result in a better GPA outcome. Another study on the relationship between physical activity and academic performance [2], however, found very small correlations between body fat/BMI and GPA ($r= 0.06$ and $r=-0.08$ respectively). Similarly, the study conducted by Gonzales et al. [5] concluded that there was no correlation between exercise levels (measured in terms of low, moderate, and high) and GPA in graduate students.

Previous research regarding the correlation between stress and students' academic success has also yielded mixed results. For instance, the StudentLife study at Dartmouth [3] concluded that "perceived stress scale negatively correlates with the spring term GPA". This study also introduced terms called "positive affect" and "negative affect", which represent how much a positive or negative mood changes across the semester. The study then noted that students with higher GPAs were generally positive but had a decreased positive affect toward the end of the semester, possibly because of the pressure coming from the finals. Another study at Universiti Putra Malaysia [4] found the same negative but weak correlation between GPA and stress with $r=-0.195$. Although the correlation between stress and grade point average is still debatable, there is evidence suggesting that a combination of depression and anxiety results in a much lower grade point average. A study by Eisenberg et al. [6] conducted at the University of Michigan reported that depression is an indicator of a low-grade point average and a higher probability of dropping out from the university.

### 2.2 Lifestyle Factors

Overall, previous research has indicated that there exists a correlation between lifestyle factors such as study habits, study duration, socialization time, and involvement within one's campus with a student's academic success. For instance, the SmartGPA [1] concludes that students who spend time at fraternities tend to do more poorly academically, but students who socialize more at their dorms are more likely to have higher GPAs. The study also correlates with increasing conversation duration with a higher GPA, which could indicate more group work and communication towards the end of the semester. Another study, conducted by DeMartini et al. [7] in the College of New Jersey, correlates higher attendance rates with a higher GPA. Using a least-squares regression and a defined probability model, this study compared the number of missing classes with GPA and concluded that students who did not miss more than three classes a week were more likely to get a GPA in the 3.1 to 4.0 range.

### 2.3 Personality Traits

Intuition about student success may suggest that students who work hard typically earn a higher grade point average. Previous research seems to support this and has found that a higher college grade point average is associated with the conscientiousness personality trait [8], which is the trait associated with being dutiful and doing one's work. Additionally, the Lasso regression analysis reported in Dartmouth's SmartGPA study [1] selected conscientiousness as the one long-term indicator of a higher grade point average. However, there is little research done to indicate how significantly one's personality impacts academic success in comparison to other environmental factors.

## 3 Methods

### 3.1 Recruitment and Data Collection

Participants in the study were from a pool of first-year undergraduate students at a Carnegie-classified R-1 University in the United States. Students were eligible to participate in the study if they were enrolled as a full-time student on campus for the semester and owned a data plan-enabled smartphone running iOS or Android. Students were invited to our lab to be screened for eligibility, provide informed consent, download a mobile application to track sensor data from their smartphones, and receive a Fitbit Flex 2 to track steps and sleep. After enrollment and at the end of the semester, the students completed online questionnaires related to their general mental health, stress, anxiety, and depression. Data was collected from smartphone and Fitbit sensors and was continuously recorded over the course of the semester (16 weeks). Out of the 188 first-year college students initially recruited, 138 completed the study and the questionnaires at the beginning and the end of the study. The questionnaires for mental health assessment were delivered via email and administered using Qualtrics – an online survey platform [9].

Students installed the AWARE framework [10] – a data collection mobile application with supporting backend and network infrastructure to collect sensor data unobtrusively from students' smartphones. This enabled us to among others record location, phone usage (i.e., when the screen status changed to on or off and locked or unlocked), and call logs for incoming, outgoing and missed calls. Further, we equipped participants with a Fitbit Flex 2, which records the number of steps and sleep status (asleep, awake, restless, or unknown). Calls and phone usage were sampled when an event such as a phone call or screen unlock took place. Location, sleep, and steps, on the other hand, were sampled at regular frequencies. In our study, location coordinates were collected every 10 minutes, sleep every minute, and steps every 5 minutes.

Data from AWARE was de-identified and automatically transferred over Wi-Fi to our backend server on a regular basis, and data from the wearable Fitbit was retrieved using the Fitbit API at the end of the study. Participants were asked to keep their phone and Fitbit charged and carry/wear them at all times. To maintain the participants' privacy and confidentiality, we stored all identifiable information (e.g., names, contact information) separate from their deidentified survey and sensor data. Only a few authorized members of the research team had access to the participants' identifiable information. The Institutional Review Board (IRB) reviewed, oversaw, and approved all procedures.

### 3.2 Feature Extraction

The AWARE data was then processed, and a number of features were extracted from location, calls, messages, phone usage, and Fitbit (incl. calories, sleep, and steps) following the procedure described in [11]. For this study, we considered features that are most relevant to the health and lifestyle categories of interest. They were also selected based on the factors that were shown to affect grade point average in previous research [1, 3, 4, 5, 6, 7, 8]. For the physical health category, we included calories burned, a number of steps taken, and time spent asleep. For lifestyle, we chose features related to the time spent at different locations such as academic facilities and outdoors as well as features that indicate movement patterns. We also considered time spent on the phone and duration of phone calls as a proxy for phone usage and online social activities. Table 1 lists the extracted phone, and Fitbit features used in our analysis. We also used questionnaire data related to mental health, including stress score, anxiety/depression score, and mental composite score as additional features.

**Table 1.** List of features extracted from the Phone and FitBit sensors we used for our analysis

| Feature Category | Feature |
|---|---|
| Physical | Calories burned, number of steps taken, time spent asleep |
| Location | Time spent at home, academic facilities, off campus, dorm apartments, social event places, green areas |
| Movement | Circadian movement, location variance, radius of gyration (the range of movement) |
| Social | Duration of phone calls |
| Phone Usage | Time spent on phone screen |

### 3.3 Model Building with Machine Learning Algorithms

To analyze the trends, the semester data was first divided into the following three periods:
- **Period 1**- the first four weeks of the semester. Since these are approximately the weeks before the extended midterm season, they are seen as weeks with a lighter course load.
- **Period 2**- the university's extended midterm season. Since there are frequent midterms, this period is seen as the most academically rigorous.
- **Period 3**- the finals week. Although there are exams, there are no classes or homework, and thus the schedule required for academic success is hypothesized to be seen as different from the other two periods.

We created data-sets that included data from each period. These data-sets were used in a classification process to label the GPA level. The classification was a binary task to output whether or not the GPA was above 3.5. This threshold was chosen based on the students' expectations of success at the target university. We built classification models of data-sets for each period using Logistic Regression (LR) and Support

Vector Machine (SVM). Both models were created using the Scikit-learn library in Python. The intuition behind our choice of algorithms was that although LR and SVM generate the same type of weight-based models, their classification strategies are sufficiently different from each other. Therefore, using two different algorithms could provide stronger evidence for the observed trends.

### 3.4    Model Interpretation

Rather than the accuracy of the classification, we were interested in interpreting the built models by the algorithms to understand and discover the behavioral features that contribute to the prediction of the GPA level. Both LR and SVM generate models that output the weights of each feature in the classification process. We take advantage of this to understand the impact of each feature on the GPA outcome. We interpret the positive and negative weights of the features as proxies for an increase or reduction in the significance of the behavioral feature in the analyzed period during the semester.

However, the models generated by LR and SVM may associate different weights to each feature based on their classification strategy. We measure a mutual agreement score based on the aggregated weight of each feature in both models. As such, the features that are heavily weighted in both models will be highlighted as dominant behavioral features.

## 4    Results

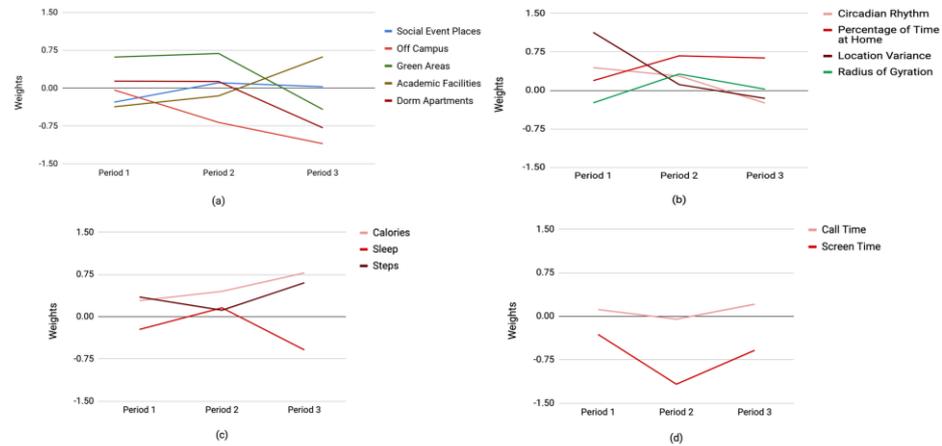

**Fig. 1.** a) The weight of location features depicts spending more time in academic facilities, and spending less time off-campus, in dorm halls, and in green areas towards the end of the semester. b) Movement features show a decrease in mobility towards the end of the semester as indicated by more time staying at home, lower location variation, and lower circadian movement. c) Physical features display almost the same direction for the number of steps and burned calories over the semester, but the amount of sleep trend is exactly in the opposite way. d) Social and phone usage features indicate the same direction for both screen and call time

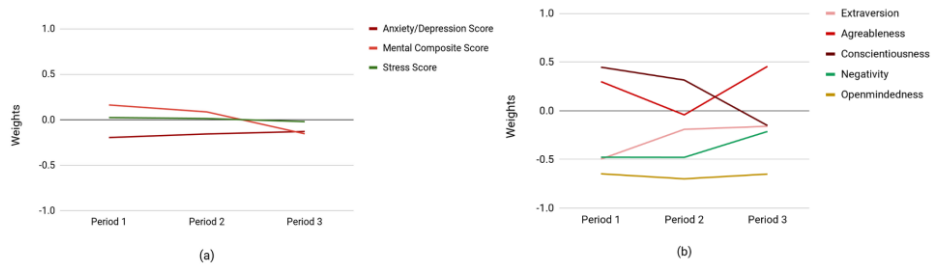

**Fig. 2.** a) Mental health features show depression/anxiety has a negative weight in the whole semester, the stress level is neutral over the semester,and general mental health score goes down towards the end of the semester. b) Personality features indicate there are significant negative weights for extraversion, negativity, and openmindedness score over the semester. However, agreeableness score has positive and increasing weight and conscientiousness weight decreases towards the end of the semester.

### 4.1　Location and Movement Patterns

The first trend that was noticed was that there were distinct trends relating to the amount of time spent at each location Figure 1(a). As the semester progressed, students with higher GPAs tended to spend time at academic facilities and the student center (where the gym is) and less time at all other places, especially off-campus and outdoors. This implies that students with a higher grade point average tended to establish a normal exercising and studying routine as their classes increased in difficulty. They also tended to socialize less as the semester progressed, instead choosing to focus on work. Another trend that is noticeable is that the weights of the features depicting how much time they spent at a certain place have the highest magnitude during finals week. This might imply that one's sensitivity in regard to where they spend their time increases by a large amount during finals week. This is also demonstrated in the movement patterns in Figure 1(b) where location variation and circadian (daily) movement decreased towards the end of the semester while the percentage of time staying at home increased in periods 2 and 3.

### 4.2　Physical Activity and Sleep Patterns

Other behavioral features associated with students' GPA were physical activity and sleep patterns, as shown in Figure 1(c). While steps and calories burned correlated in the same direction with a decrease in steps in period 2and an increase in period 3, the sleep duration showed an opposite pattern with an increase in period 2 and a sharp decrease in period 3. Although period 2 in our categorization is mainly midterm period, it also includes the spring break where students often catch up with sleep. That may explain the positive weight of sleep in period 2. In period3, the lack of sleep is evident as the students may stay up late for their final preparation.

### 4.3 Social and Phone Usage Patterns

Analysis of social-related factors, such as call time Figure 1 (d) indicates that although the call activities may slightly decrease as the workload increases in period 2, there is a rather stable pattern of call communication in students with higher GPA in all three periods with a slightly higher weight in period 3. An explanation of this phenomenon could be that as the workload increases during period 2, students respond to the extra schoolwork by directing time toward academics. However, these students know that they need to relax during finals week, relying on social support to help them through finals. The negative weight of screen time in all three periods may indicate that the amount of time on the phone is generally lower in highly successful students compared to students with lower GPAs and that as the pressure of the school work increases, successful students spend less time on their phones.

### 4.4 Mental Health Factors

We also built models with mental health scores acquired from the questionnaires for each period. As shown in Figure 2, (a) while depression and anxiety have an almost constant negative weight in all three periods, the perceived stress level remains neutral, and the general mental health score contributes positive weights in period 1 and 2 and as lightly negative weight in period 3. Although the neutral weight of stress may be counter-intuitive, it may suggest that the perceived stress level has no effect on the student's success. The constant weight of anxiety and depression scores confirms that depression as a longitudinal mental health problem has negative effects on an individual's performance.

### 4.5 Personality Factors

When looking at the weights of each personality trait, it is noted that there are significant negative weights as-sociated with the personality traits of extraversion, negativity, and openmindedness in all three periods with slight decreases in period 3 for negativity and extraversion. The negative weight of extraversion in predicting GPA may be because extraversion is an indicator of other lifestyle behaviors that contribute to a lower grade point average, such as extracurricular or off-campus activities and spending less time on coursework. Meanwhile, the negativity pattern may also indicate its impact on one's focus, which can impact one's final grades. On the other hand, the constant negative weight of openmindedness in all three periods suggests that since these students are more willing to try new things, they may have a more unstable schedule, which can contribute to a more disorganized lifestyle and, therefore, poor academic performance. The increasing positive weights of agreeableness score in period 3 may indicate better success because of better teamwork that more agreeable people are capable of doing. In conclusion, the findings have shown that instead of certain personality traits, ensuring that one does well academically, there exist personality traits ensuring that one does poorly.

### 4.6 Comparison with Previous Work

Similar to previous work by the Dartmouth team [5, 1], we found a decrease in mobility towards the end of the semester indicated by more time staying at home, lower location variation, lower circadian movement, and smaller range of movement. Our study is the first to identify clear trends in the locations that students spend time in during different periods, including more time in academic facilities and less time off-campus and outdoors during finals. Our study also shows clear trends in the impact of physical activity and sleep on predicting GPA in each period. While the Dartmouth team only looked at the trends of activities over the semester, the trends in both studies seem to agree, i.e., more successful students have a higher level of physical activities, especially during finals weeks. Our study, however, shows the trends in sleep data over the three periods as well, which has not been discussed in previous work.

The partying factor that was clearly a contributing factor to GPA prediction does not seem evident in our results. One reason may be that our only proxy of partying is the time spent at social event places on-campus, which may not be enough compared to collecting self-report data about partying in the Dartmouth study. We, however, see significant negative weights in screen time for predicting GPA, indicating potential control over phone use in successful students. This trend was not evident in previous research.

The surprising difference in our results comes from the fact that perceived stress has a neutral weight in GPA prediction, which is the opposite of Dartmouth's findings [5, 1]. While more analyses are needed to replicate our results, we argue that although the stress level among college students in top tier universities is generally high, it may be a driving factor for them rather than negatively affecting their academic performance. Our observations regarding the impact of depression and anxiety, however, is consistent with all studies we found including [6].

The impact of personality factors is somewhat similar to previous work in that conscientiousness has a positive weight in predicting the GPA for the majority of the semester time. However, we also observe clear trends in the positive weight of agreeableness in student performance that may be a proxy of successful teamwork. We believe the teamwork trend is observed in the SmartGPA study [1] in the form of increased conversations at the end of the semester. The more students work together, the more conversations they tend to have.

## 5   Conclusion

Our analysis of smartphone, Fitbit, and questionnaire data collected from 138 students over the course of a semester revealed health and behavioral trends associated with students' academic performance measured by the grade points average (GPA). We built machine learning models of the data for three distinct periods during the semester and used the parameter weights of those models to interpret the effect of behavioral features in predicting the GPA. Our results showed increased time at home and academic facilities, decreased location variation and movement (a proxy of more study time), and increased physical activity towards the finals week in students with higher GPAs. We also observed less phone usage and moderate social activity in such students. While we found no effect of perceived stress on GPA, we observed a constant negative weight of depression and anxiety. Most of our observations are aligned with the previous work in this domain. However, more analyses are to be done to replicate these results on the impact of health and behavioral factors on students' performance.